\documentclass[aps,prd,onecolumn,groupedaddress,showpacs,nofootinbib,amssymb]{revtex4}
\usepackage[dvips]{graphicx}
\usepackage{amssymb}
\usepackage{amsmath}
\usepackage{graphicx,,color}
\usepackage{amsfonts}
\usepackage{bm}
\usepackage{cancel}
\usepackage{comment}
\usepackage[dvipsnames]{xcolor}
\usepackage{makecell}

\allowdisplaybreaks[4]

\begin{document}

\tolerance=5000

\title{$\mathcal{R}^2$ Quantum Corrected Scalar Field Inflation}
\author{V.K. Oikonomou,$^{1,2}$}\email{v.k.oikonomou1979@gmail.com,voikonomou@auth.gr}
\author{Ifigeneia Giannakoudi,$^{1}$}\email{ifigeneiagiannakoudi@gmail.com,ifgianna@auth.gr}
\affiliation{$^{1)}$Department of Physics, Aristotle University of Thessaloniki, Thessaloniki 54124, Greece\\
$^{2)}$ Laboratory for Theoretical Cosmology, Tomsk State
University of Control Systems and Radioelectronics  (TUSUR),
634050 Tomsk, Russia}

\tolerance=5000

\begin{abstract}
String theory enjoys an elevated role among quantum gravity
theories, since it seems to be the most consistent UV completion
of general relativity and the Standard Model. However, it is hard
to verify the existence of this underlying theory on terrestrial
accelerators. One way to probe string theory is to study its
imprints on the low-energy effective inflationary Lagrangian,
which are quantified in terms of high energy correction terms. It
is highly likely, thus, to find higher order curvature terms
combined with string moduli, that is scalar fields, since both
these types of interactions and matter fields appear in string
theory. In this work we aim to stress the probability that the
inflationary dynamics are controlled by the synergy of scalar
fields and higher order curvature terms. Specifically, we shall
consider a well motivated quantum corrected canonical scalar field
theory, with the quantum corrections being of the $\mathcal{R}^2$
type. The reason for choosing minimally coupled scalar theory is
basically because if scalar fields are evaluated in their vacuum
configuration, they will either be minimally coupled or
conformally coupled. Here we choose the former case, and the whole
study shall be performed in the string frame (Jordan frame), in
contrast to similar studies in the literature where the Einstein
frame two scalar theory is considered. We derive the field
equations of the quantum-corrected theory at leading order and we
present the form the slow-roll indices obtain for the quantum
corrected theory. We exemplify our theoretical framework by using
the quadratic inflation model, and as we show, the $\mathcal{R}^2$
quantum corrected quadratic inflation model produces a viable
inflationary phenomenology, in contrast with the simple quadratic
inflation model.
\end{abstract}

\maketitle

\section{Introduction}

String theory is to date the most successful candidate for
describing the pre-inflationary quantum state of the Universe,
since it unifies gravity with the Standard Model fundamental
interactions in a consistent way. However, the drawback of this
theory is that the predictions of it cannot be tested for the
moment on terrestrial particle accelerators. The LHC has achieved
a 15 TeV center of mass energy scale and no sign of alternative to
the Standard Model physics seems to exist. Even supersymmetry does
not seem to exist for scales up to 15 TeV center of mass thus, for
the moment, high energy physics are not directly linked to the
world we can access experimentally. This however is not a problem
of string theory, it is our problem since we need to find
alternative ways to test high energy physics realistically in the
world we access. One consistent and elegant way to obtain hints on
high energy physics works, is to try to find traces of the high
energy physics Lagrangian, to the low energy effective Lagrangian
which describes the Universe in the early stages of its classical
evolution. To date, it is believed that the Universe after the
quantum epoch entered the inflationary era
\cite{inflation1,inflation2,inflation3,inflation4}, during which
the Universe is theorized to be classical and four dimensional.
But as we mentioned, it is quite possible to find high energy
physics traces in the inflationary Lagrangian, in the form of
quantum corrections. This possibility is sensational since it will
provide insights that, for the moment, are hard to obtain from
terrestrial accelerators. Inflation and early-time cosmology will
be tested in the following two decades, and the scientific
community hopes that some light will eventually be shed on the
early-time physics and strong answers will be given on the
question of how our Universe works. There are two ways to test the
early-time era, and both of these aim to the sky and do not rely
on terrestrial particle accelerators. The first way is to seek
traces of inflation originating patterns in the Cosmic Microwave
Background (CMB) temperature fluctuations. The future experiments
which are basically the fourth generation CMB experiments
\cite{CMB-S4:2016ple,SimonsObservatory:2019qwx}, will seek for the
$B$-modes of inflation (curl mode) \cite{Kamionkowski:2015yta}. It
is known that $B$-modes can be generated in two ways, from the
gravitational lensing generated transition of $E$-modes to
$B$-modes at small angular scales or equivalently large multipoles
of the CMB, or from primordial tensor modes, the so called
primordial gravitational waves, at large angular scales or
equivalently low multipoles of the CMB. Thus the CMB-based way to
detect primordial gravitational wave generated patterns on the
CMB, is to look at the low multipoles of the CMB. The other way to
probe the post-quantum era of our Universe, is more direct than
the CMB way, and it will be based on future space high frequency
interferometers
\cite{Hild:2010id,Baker:2019nia,Smith:2019wny,Crowder:2005nr,Smith:2016jqs,Seto:2001qf,Kawamura:2020pcg,Bull:2018lat}.
These interferometers will seek directly for a stochastic
background of primordial gravitational waves. This observation
will be sensational to say the least, and will stir things up
significantly in both high energy physics and theoretical
cosmology.

Most of the inflationary theories use the single scalar field
description, which is simple to tackle analytically. But as we
mentioned, it is highly likely that the string theory era will
have its effect on the low-energy effective inflationary
Lagrangian. These low-energy string oriented effects come in terms
of higher order curvature corrections \cite{Codello:2015mba}.
Therefore, it is possible that not just a single scalar field
controls the dynamics of inflation, but the synergy of a scalar
field and the higher order curvature terms as part of some
modified gravity terms
\cite{reviews1,reviews2,reviews3,reviews4,reviews5,reviews6},
might actually control the inflationary era. This issue started to
appear in the recent literature of theoretical cosmology
\cite{Ema:2017rqn,Ema:2020evi,Ivanov:2021ily,Gottlober:1993hp,delaCruz-Dombriz:2016bjj,Enckell:2018uic,Karam:2018mft,Kubo:2020fdd,Gorbunov:2018llf,Calmet:2016fsr,Oikonomou:2021msx},
and it is not simply an academic exercise, but might be a
compelling task in the future. The motivation for using combined
scalar field and higher order gravity Lagrangians is apparent,
both scalar fields (string moduli) and higher order gravity terms
originate from the underlying string theory. Thus, one must
include both these ingredients in the low-energy Lagrangian in
order to be accurate and scientifically ''democratic'' towards
treating the higher order fields and terms. Single field
approaches and pure modified gravity terms are simpler to treat
analytically, but this is not a compelling motivation to exclude
the probability of finding combined effects in the inflationary
Lagrangian. Another motivation comes from the fact that
non-tachyon single scalar field theories and fundamental modified
gravity theories, like for example $f(\mathcal{R})$ gravity, seem
to predict a significantly low energy spectrum of primordial
gravitational waves. Thus, if a future direct signal of stochastic
gravitational waves is detected, this will certainly exclude these
descriptions, unless some exotic scenario occurs, which however
must enhance the spectrum $10^4-10^8$ times in order to be
detected. Thus combinations of single scalar field descriptions
with modified gravity terms seem to be an elegant possibility,
since in these theories, significant enhancement of the primordial
gravitational wave energy spectrum might occur, see for example
Ref. \cite{Odintsov:2021kup}. When the scalar field is evaluated
in its vacuum configuration, it either is minimally coupled or
conformally coupled, and thus the quantum corrections will either
appear in conformally coupled single scalar field Lagrangians, or
minimally coupled scalar field Lagrangians. In most works
appearing to date in the literature, the first perspective is
studied, where the inflationary phenomenology of the Higgs scalar
is considered in the presence of $\mathcal{R}^2$ quantum
corrections
\cite{Ema:2017rqn,Ema:2020evi,Ivanov:2021ily,Gottlober:1993hp,delaCruz-Dombriz:2016bjj,Enckell:2018uic,Karam:2018mft,Kubo:2020fdd,Gorbunov:2018llf,Calmet:2016fsr}.
However, all the existing approaches study the inflationary
phenomenology in the Einstein frame, thus the resulting theory has
effectively two scalar fields. There is another possibility
though, to investigate the phenomenology of these combined scalar
field-modified gravity theories in the string frame (Jordan
frame). This perspective was investigated for power-law
$f(\mathcal{R})\sim \mathcal{R}^n$ gravity in
\cite{Oikonomou:2021msx}, however the most important quantum
correction, the $\mathcal{R}^2$ was not studied, since it could
not be derived formally for the general $n$ case. This is due to
the fact that for $n=2$ the situation is much more simple and more
easy to tackle analytically, than the general $\mathcal{R}^n$
case. In this paper we shall study minimally coupled single scalar
field theory in the presence of $\mathcal{R}^2$ quantum
corrections in the string frame. We shall derive the field
equations, and by using the slow-roll assumptions, and physically
motivated assumptions, we shall quantify the effect of the
$\mathcal{R}^2$ quantum corrections on single scalar field
inflation. The resulting effective field equations acquire an
elegant final form as we will demonstrate, and accordingly the
inflationary phenomenology of the resulting theory will thoroughly
be investigated.

This work is organized as follows: In section II we shall present
the general $\mathcal{R}^2$ quantum corrected single field
Lagrangian and field equations. We quantify the leading order
$\mathcal{R}^2$ quantum corrections in the field equations, and we
express the slow-roll indices and the corresponding observational
indices in terms of the scalar field, in which the $\mathcal{R}^2$
quantum effects are included. In section III we analyze in depth
one example model which yields a viable inflationary era
compatible with the latest CMB constraints on inflation. Finally,
the conclusions follow in the end of the paper.

\section{Canonical Scalar Field Inflation in the presence of $R^2$ Gravity: Formalism}

The most general four dimensional scalar field Lagrangian
containing at most two derivatives is,
\begin{equation}\label{generalscalarfieldaction}
\mathcal{S}_{\varphi}=\int
\mathrm{d}^4x\sqrt{-g}\left(\frac{1}{2}Z(\varphi)g^{\mu
\nu}\partial_{\mu}\varphi
\partial_{\nu}\varphi+\mathcal{V}(\varphi)+h(\varphi)\mathcal{R}
\right)\, ,
\end{equation}
and when the scalar fields are evaluated at their vacuum
configuration, the scalar field must be either minimally coupled
of conformally coupled. We shall focus on the first possibility in
which $Z(\varphi)=-1$ and $h(\varphi)=1$ in the action
(\ref{generalscalarfieldaction}). Now the quantum corrections
local effective action which is consistent with diffeomorphism
invariance and contains up to fourth order derivatives, is
\cite{Codello:2015mba},
\begin{align}\label{quantumaction}
&\mathcal{S}_{eff}=\int
\mathrm{d}^4x\sqrt{-g}\Big{(}\Lambda_1+\Lambda_2
\mathcal{R}+\Lambda_3\mathcal{R}^2+\Lambda_4 \mathcal{R}_{\mu
\nu}\mathcal{R}^{\mu \nu}+\Lambda_5 \mathcal{R}_{\mu \nu \alpha
\beta}\mathcal{R}^{\mu \nu \alpha \beta}+\Lambda_6 \square
\mathcal{R}\\ \notag &
+\Lambda_7\mathcal{R}\square\mathcal{R}+\Lambda_8 \mathcal{R}_{\mu
\nu}\square \mathcal{R}^{\mu
\nu}+\Lambda_9\mathcal{R}^3+\mathcal{O}(\partial^8)+...\Big{)}\, ,
\end{align}
where the parameters $\Lambda_i$, $i=1,2,...,6$ are appropriate
dimensionful constants. Also non-analytical ``leading logs''
correction terms of the form $\mathcal{R}^2 \log \mathcal{R}$
might be present and also non-local terms $\mathcal{R} \log
\frac{-\square \mathcal{R}}{\mu^2}$, the phenomenology of which
was analyzed in \cite{Odintsov:2017hbk}. In this paper, we shall
focus on the $\mathcal{R}^2$ corrections, which is among the
simplest corrections one can consider. Thus, the minimally coupled
single scalar field action including the $\mathcal{R}^2$
corrections is effectively an $f(\mathcal{R},\varphi)$ action of
the form,
\begin{equation}
\label{action} \centering
\mathcal{S}=\int{d^4x\sqrt{-g}\left(\frac{
f(\mathcal{R})}{2\kappa^2}-\frac{1}{2}g^{\mu
\nu}\partial_{\mu}\varphi
\partial_{\nu}\varphi-\mathcal{V}(\varphi)\right)}\, ,
\end{equation}
where $\kappa^2=8\pi G=\frac{1}{Mp^2}$ and $M_p$ stands for the
reduced Planck mass, and also,
\begin{equation}\label{fr}
    f(\mathcal{R})=\mathcal{R}+\frac{\mathcal{R}^2}{36M^2}\, ,
\end{equation}
with $M$ being a mass scale to be determined later on. Note that,
since the model we are considering is an $f(\mathcal{R},\varphi)$
theory in the string frame, the value of $M$ is not required to be
that of the standard $\mathcal{R}^2$ model
\cite{Starobinsky:1980te,Bezrukov:2007ep}, and it will generally
take different values from the usual ones obtained by
phenomenological reasons for the vacuum $\mathcal{R}^2$ model in
the Einstein frame \cite{Appleby:2009uf}. For the background
metric, we assume a flat Friedmann-Robertson-Walker (FRW) metric
with line element,
\begin{equation}
    \centering\label{metric}
    \mathrm{d}s^2 = - \mathrm{d} t^2 + a(t) \sum_{i = 1}^3 \mathrm{d} x_i^2,
\end{equation}
with $\alpha(t)$ being the scale factor and the corresponding
Hubble rate is $\mathcal{H}=\frac{\dot{a}}{a}$. Upon varying the
gravitational action (\ref{action}) with respect to the metric, we
obtain the following field equations,
\begin{equation} \label{Friedmann}
    3 f_{\mathcal{R}} \mathcal{H}^2=\frac{R f_{\mathcal{R}} -f}{2}-3\mathcal{H} \dot F_{\mathcal{R}}+\kappa^2\big(\frac{1}{2}\dot\varphi^2+\mathcal{V}(\varphi)\big) \ ,
\end{equation}
\begin{equation} \label{Raychad}
    -2 f_{\mathcal{R}} \dot{\mathcal{H}} = \kappa^2 \dot\varphi^2 + \ddot f_{\mathcal{R}} -\mathcal{H} \dot f_{\mathcal{R}} \ ,
\end{equation}
\begin{equation} \label{fieldeqmotion}
   \ddot\varphi+3\mathcal{H}\dot\varphi+\mathcal{V}'=0 \ ,
\end{equation}
where the ``dot'' denotes the derivative with respect to cosmic
time, the ``prime'' describes derivative with respect to the
scalar field $\varphi$ and $f_{\mathcal{R}}=\frac{\partial
f}{\partial \mathcal{R}}$. Taking into consideration that
$f(\mathcal{R})=\mathcal{R}+\frac{\mathcal{R}^2}{36M^2}$ and also
that the Ricci scalar and its derivative for the FRW are given by,
\begin{equation}\label{ricciscalar}
    \mathcal{R} =12 \mathcal{H}^2 + 6\dot{\mathcal{H}} \, , \ \dot{\mathcal{R}}=24\mathcal{H}\dot{\mathcal{H}} +6\ddot{\mathcal{H}},
\end{equation}
the Friedmann and Raychaudhuri equations become,
\begin{equation} \label{Friedman2}
    3\mathcal{H}^2+\frac{3}{M^2}\mathcal{H}^2 \dot{\mathcal{H}}= \frac{\dot{\mathcal{H}}^2}{2}-\frac{\ddot{\mathcal{H}} \mathcal{H}}{M^2}+\kappa^2\big(\frac{1}{2}\dot \varphi^2 +\mathcal{V}(\varphi) \big) ,
\end{equation}
\begin{equation} \label{Raycha2}
    -2 \dot{\mathcal{H}}-\frac{2}{M^2}\dot{\mathcal{H}}^2 = -\frac{\ddot{\mathcal{H}} \mathcal{H}}{M^2}+\kappa^2 \dot \varphi^2 ,
\end{equation}
with a remarkable cancellation of the terms $\mathcal{H}^2
\dot{\mathcal{H}}$ taking place in the latter one, which is absent
in the case that a power-law $\mathcal{R}^n$ gravity is
considered, with $n \neq 2$, see \cite{Oikonomou:2021msx}.  The
slow-roll conditions during the inflationary era demand that,
\begin{equation}\label{slowrollH}
    \dot{\mathcal{H}} \ll \mathcal{H}^2 \ , \ \ddot{\mathcal{H}} \ll \mathcal{H} \dot{\mathcal{H}},
\end{equation}
however the term $\frac{\mathcal{H}^2 \dot{\mathcal{H}}}{M^2}$ is
possibly of the same order as $\mathcal{H}^2$, at least for a
quasi-de Sitter evolution. Thus, we can easily neglect the terms
containing the higher order Hubble rate derivative $\sim
\ddot{\mathcal{H}}$ and we shall also assume that the following
approximations hold true,
\begin{equation}\label{approxH}
    \frac{\dot{\mathcal{H}}^2}{M^2} \ll \mathcal{H}^2,\,\,\,\frac{\dot{\mathcal{H}}^2}{M^2} \ll
    \mathcal{V}(\varphi)\, .
\end{equation}
Although the above approximations are well justified for a usual
quasi-de Sitter evolution, it is compelling to examine that these
hold true for a viable inflationary phenomenology, for the same
values of the free parameters which guarantee the viability of the
model. Moreover, in the same fashion as single scalar field
inflation, we will require the scalar field to satisfy,
\begin{equation}\label{approxphi}
   \frac{1}{2} \dot \varphi^2 \ll \mathcal{V}(\varphi) \ , \ \ddot \varphi \ll \mathcal{H}\dot
   \varphi .
\end{equation}
In effect of all the above, the field equations take the form,
\begin{equation} \label{Friedman3}
    3\mathcal{H}^2+\frac{3\mathcal{H}^2}{M^2} \dot{\mathcal{H}}\simeq  \kappa^2 \mathcal{V}(\varphi),
\end{equation}
\begin{equation} \label{Raycha3}
     -2 \dot{\mathcal{H}}-\frac{2}{M^2}\dot{\mathcal{H}}^2 \simeq \kappa^2 \dot \varphi^2  ,
\end{equation}
\begin{equation} \label{dotphi}
   \dot \varphi \simeq - \frac{\mathcal{V}'}{3\mathcal{H}} .
\end{equation}
We notice that the Raychaudhuri equation is practically a second
order polynomial equation with respect to $\dot{\mathcal{H}}$,
thus the solution is,
\begin{equation}\label{dothanalyticsol}
\dot{\mathcal{H}}=\frac{-M^2+M\sqrt{M^2-2\dot{\varphi}^2\kappa^2}}{2}\,
,
\end{equation}
and the Friedmann equation becomes,
\begin{equation}\label{friedaux}
\frac{3\mathcal{H}^2}{2}-\frac{3\mathcal{H}^2\sqrt{M^2-2\dot{\varphi}^2\kappa^2}}{2M}\simeq
\kappa^2 \mathcal{V}(\varphi)\, .
\end{equation}
In the following we shall use the approximation,
\begin{equation}\label{taylorphi}
    \frac{2\kappa^2 \dot \varphi^2}{M^2} \ll 1\, ,
\end{equation}
The validity of the approximations (\ref{approxH}) and
(\ref{taylorphi}) must be tested for any potential viable
inflationary model. As long as the above approximation holds true,
the Friedmann and Raychaudhuri equations acquire the following
form at leading order,
\begin{equation} \label{Friedman}
    \mathcal{H}^2 \simeq \frac{\kappa^2 \mathcal{V}(\varphi)}{3}+\mathcal{O}(\frac{\kappa^2 \dot{\varphi}^2}{2}\mathcal{H}^2),
\end{equation}
\begin{equation} \label{Raycha}
     \dot{\mathcal{H}} \simeq -\frac{\kappa^2 \dot{\varphi}^2}{2}  -\frac{\kappa^4
     \dot{\varphi}^4}{4M^2}\, .
\end{equation}
The resulting equations of motion, namely Eqs. (\ref{Friedman}),
(\ref{Raycha}) and the slow-roll scalar field equation, namely Eq.
(\ref{dotphi}) will be the starting point of our analysis.
Basically these equations depict the direct effect of the $R^2$
corrections on the standard scalar field inflation in the Jordan
frame. The second term in the Raychaudhuri equation is basically a
quantum correction in the field equations introduced by the
presence of the $\mathcal{R}^2$ quantum corrections in the
effective inflationary Lagrangian, when of course the condition
(\ref{taylorphi}) holds true.  The effects of the $\mathcal{R}^2$
gravity term are quantified in the Raychaudhuri equation
(\ref{Raycha}), and will also affect crucially the slow-roll
indices. The slow-roll indices for a general
$f(\mathcal{R},\varphi)$ theory are defined \cite{Hwang:2005hb},
\begin{align}\label{slowrollindices}
\epsilon_1=-\frac{\dot{\mathcal{H}}}{\mathcal{H}^2},\,\,\,\epsilon_2=\frac{\ddot
\varphi }{\mathcal{H} \dot \varphi},\,\,\,
\epsilon_3=\frac{\dot f_{\mathcal{R}} }{2\mathcal{H}f_{\mathcal{R}}},\,\,\,\epsilon_4=\frac{\dot E}{2\mathcal{H}E}\, ,\\
\end{align}
where,
\begin{equation}\label{E1}
    E=f_{\mathcal{R}}+\frac{3 \dot f_{\mathcal{R}}^2}{3\kappa^2\dot
    \varphi^2}\, .
\end{equation}
Using the equations (\ref{Friedman}) and (\ref{Raycha}), the first
slow-roll index $\epsilon_1$ is easily found equal to,
\begin{equation}\label{e1}
    \epsilon_1=\frac{1}{2\kappa^2} \left( \left(\frac{\mathcal{V}'}{V}\right)^2 +\frac{1}{6M^2}{\left(\frac{\mathcal{V}'}{V}\right)}^2\frac{\mathcal{V}'^2}{V}\right)\, .
\end{equation}
As for the expression of $\epsilon_2$ we also use (\ref{dotphi}),
from which we calculate $\ddot{\varphi}$, so after some algebra,
the second slow-roll index reads,
\begin{equation}\label{e2}
    \epsilon_2=-\frac{\mathcal{V}''}{\kappa^2 V}+\epsilon_1 \, .
\end{equation}
Also the resulting expression for the slow-roll index $\epsilon_3$
yields,
\begin{equation}\label{e3}
    \epsilon_3=\frac{\epsilon_1}{-1-\frac{3M^2}{2\mathcal{H}^2}+\frac{\epsilon_1}{2}} \, .
\end{equation}
The expression of $\epsilon_4$ is the most complex one so far. We
firstly calculate $E$ and $\dot E$,
\begin{equation}\label{E}
    E=1+\frac{2\mathcal{R}}{36M^2}+\frac{8}{3\kappa^2 M^4}\frac{\mathcal{H}^2 \dot{\mathcal{H}}^2}{\dot \varphi^2} ,
\end{equation}
\begin{equation}\label{dotE}
    \dot E=\frac{4 \mathcal{H} \dot{\mathcal{H}}}{3M^2} + \frac{16}{3 \kappa^2 M^4 \dot \varphi^4}\big( \mathcal{H} \dot{\mathcal{H}}^3 \dot \varphi^2-\mathcal{H}^2 \dot{\mathcal{H}}^2 \dot \varphi \ddot \varphi \big),
\end{equation}
thus, the full expression of $\epsilon_4$ can be obtained by
substituting (\ref{E}), (\ref{dotE}) to $\epsilon_4$ in
(\ref{slowrollindices}). Therefore, the dynamical evolution of the
inflationary era for the $\mathcal{R}^2$ quantum corrected
minimally coupled canonical scalar field theory is described by
Eqs. (\ref{e1})-(\ref{dotE}), in which the effects of the
$\mathcal{R}^2$ terms are clearly seen. This will be the starting
point for studying the inflationary phenomenology of a scalar
field model with an arbitrary scalar potential.

Let us move on to the calculation of the spectral index of the
primordial scalar curvature perturbations, the tensor-to-scalar
ratio and the tensor spectral index. In the case of a slow-roll
expansion, for which the slow-roll indices satisfy $\dot
\epsilon_i \ll 1$, $i=1,2,3,4$, the scalar spectral index is
written in terms of the slow-roll indices as \cite{Hwang:2005hb},
\begin{equation}\label{ns}
    n_\mathcal{S}= 1 - \frac{4\epsilon_1 + 2\epsilon_2 -2 \epsilon_3 + 2 \epsilon_4}{
 1 -\epsilon_1} .
\end{equation}
Now, the ratio of the tensor power spectrum over the scalar one
for a general $f(R,\varphi)$ theory is given by
\cite{Hwang:2005hb},
\begin{equation}\label{r}
    r=16(\epsilon_1 + \epsilon_3) ,
\end{equation}
and this is the one we will use in our analysis hereafter.
Finally, the tensor spectral index is,
\begin{equation}\label{tensorspectral}
n_{\mathcal{T}}\sim -2(\epsilon_1+\epsilon_3)\, .
\end{equation}
Using Eqs. (\ref{ns})-(\ref{tensorspectral}), the phenomenology of
an arbitrary model can directly be tested. Finally, the expression
that gives the e-foldings number is,
\begin{equation}\label{N1}
    N = \int_{t_i} ^{t_f} \mathcal{H} d t=\int_{\varphi_i} ^{\varphi_f} \frac{\mathcal{H}}{\dot \varphi} d \varphi ,
\end{equation}
where $t_i$ and $\varphi_i$ are the time instance marking the
beginning of the inflation and the value of the scalar field at
this point respectively, while $t_f$ and $\varphi_f$ are the
corresponding values at the end of the inflationary era. Using
Eqs. (\ref{Friedman}) and (\ref{dotphi}), the $e$-foldings number
takes the following form,
\begin{equation}\label{N}
    N=\int_{\varphi_f} ^{\varphi_i} \kappa^2 \frac{V}{\mathcal{V}'} d \varphi .
\end{equation}
Another important ingredient of a viable inflationary
phenomenology is related to the amplitude of scalar perturbations
$\mathcal{P}_{\zeta}(k_*)$, which is defined as follows,
\begin{equation}\label{definitionofscalaramplitude}
\mathcal{P}_{\zeta}(k_*)=\frac{k_*^3}{2\pi^2}P_{\zeta}(k_*)\, ,
\end{equation}
and should be evaluated at the pivot scale $k_*=0.05$Mpc$^{-1}$,
relevant for CMB observations. Since basically we are considering
an $f(\mathcal{R},\varphi)$ theory, the free parameters of the
theory can and should be different from the single scalar and
vacuum $\mathcal{R}^2$ gravity cases. The amplitude of the scalar
perturbations is constrained by the Planck data as follows:
$\mathcal{P}_{\zeta}(k_*)=2.196^{+0.051}_{-0.06}$. For the
generalized $f(\mathcal{R},\varphi)$ gravity, the amplitude of the
scalar perturbations in the slow-roll approximation reads
\cite{Hwang:2005hb},
\begin{equation}\label{powerspectrumscalaramplitude}
\mathcal{P}_{\zeta}(k)=\left(\frac{k \left((-2
\epsilon_1-\epsilon_2-\epsilon_4) \left(0.57\, +\log \left(\left|
\frac{1}{1-\epsilon_1}\right| \right)-2+\log
(2)\right)-\epsilon_1+1\right)}{2 \pi  z}\right)^2\, ,
\end{equation}
where $z=\frac{(\dot{\varphi} k) \sqrt{\frac{E(\varphi
)}{f_{\mathcal{R}/\kappa^2 }}}}{\mathcal{H}^2 (\epsilon_3+1)}$,
and note that at horizon crossing we have $k=a\mathcal{H}$ and the
conformal time at the horizon crossing is
$\eta=-\frac{1}{a\mathcal{H}}\frac{1}{-\epsilon_1+1}$. Thus, the
amplitude of the scalar perturbations shall be evaluated at the
first horizon crossing, so for $k_*=a\mathcal{H}$. Also note that,
a viable set of values for the spectral index, the
tensor-to-scalar ratio and the tensor spectral index, must also be
accompanied by a viable value for the amplitude of the scalar
perturbations. Thus the free parameters of the combined
$\mathcal{R}^2$ corrected theory are very much constrained. A
similar expression for the amplitude of the scalar perturbations
is,
\begin{equation}\label{ps}
    \mathcal{P}_{\zeta}(k)=\left(\frac{\mathcal{H}}{2\pi z}\left(1-\epsilon_1+\left(-2\epsilon_1 -\epsilon_2 + \epsilon_3-\epsilon_4\right)\left(\ln{\frac{1}{1-\epsilon_1}}-2+\ln{2}+0.57\right)\right)\right)^2 ,
\end{equation}
but eventually, the expressions
(\ref{powerspectrumscalaramplitude}) and (\ref{ps}), yield the
same results, so using one of the two suffices.

Summarizing, the phenomenology of an arbitrary scalar potential in
the framework of this $f(\mathcal{R})$ gravity can be studied by
using Eqs. (\ref{e1}), (\ref{e2}), (\ref{e3}), (\ref{ns}),
(\ref{r}), (\ref{ps}), (\ref{N}). Beginning with obtaining
$\varphi_f$ by solving the equation
$\epsilon_1(\varphi_f)=\mathcal{O}(1)$ and solving (\ref{N})
analytically with respect to $\varphi_i$, one can calculate the
slow-roll indices for $\varphi=\varphi_i$ and, therefore, the
scalar and tensor spectral indices and the tensor-to-scalar ratio
as functions of the model's free parameters and the $e$-foldings
number. Those steps are followed in the next section for a simple
power-law potential, the quadratic inflation potential, to study
whether a viable inflationary phenomenology can occur.


\section{Application of the Formalism: $R^2$ corrected Quadratic Inflation Potential}

In this section we shall apply the framework developed in the previous section in the case of a simple scalar power-law potential of the following form,
\begin{equation}\label{V}
    \mathcal{V}(\varphi)=\frac{\mathcal{V}_0}{\kappa^4}(\kappa \varphi)^2,
\end{equation}
where $\mathcal{V}_0$ is a dimensionless free parameter. The
quadratic inflation model is not a viable model when it is
considered alone without the presence of the $\mathcal{R}^2$
correction term, however as we now demonstrate, the
$\mathcal{R}^2$ term remedies the non-viability of the model. We
begin by substituting this potential in Eqs. (\ref{e1}),
(\ref{e2}), (\ref{e3}),(\ref{ns}), (\ref{r}), (\ref{ps}),
(\ref{N}) and we move on to finding $\varphi_f$ by solving the
equation $\epsilon_1(\varphi_f)=1$. For this potential,
$\epsilon_1$ takes a very simple form, which is,
\begin{equation}\label{e1V}
    \epsilon_1=\frac{\frac{4}{\varphi^2}+\frac{8 \mathcal{V}_0}{3M^2\kappa^2
    \varphi^2}}{2\kappa^2}\, ,
\end{equation}
thus solving the equation $\epsilon_1\simeq \mathcal{O}(1)$ we
obtain easily the final value of the scalar field at the end of
inflation, which is,
$\varphi_f=\frac{\sqrt{\frac{2}{3}}\sqrt{2\mathcal{V}_0+3M^2\kappa^2}}{M\kappa^2}$.
\begin{figure}
\centering
\includegraphics[width=18pc]{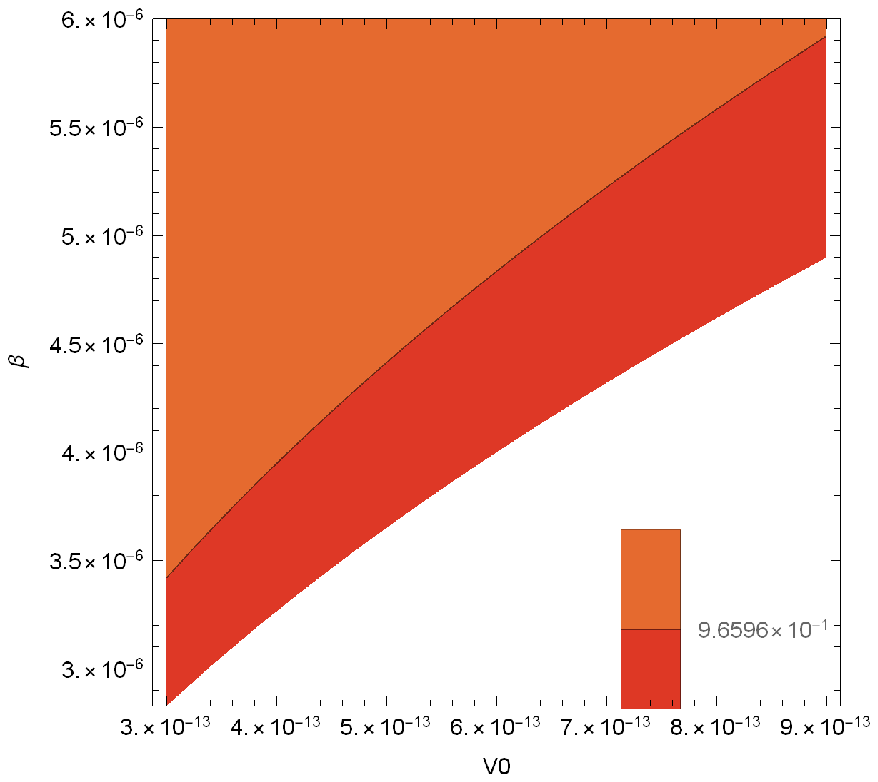}
\includegraphics[width=18pc]{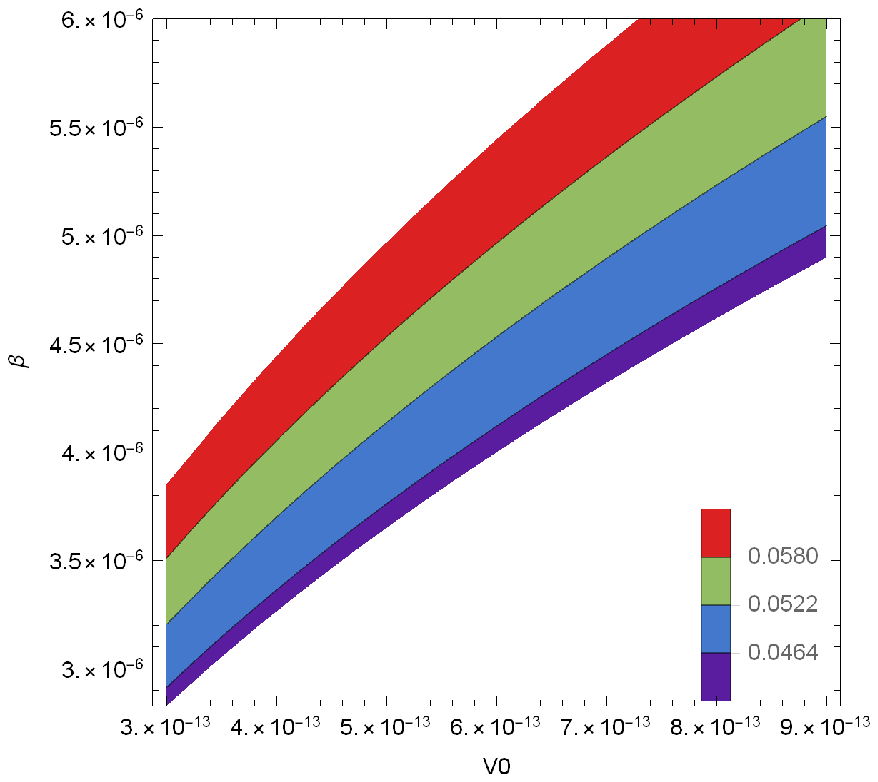}
\caption{Contour plots of the spectral index of primordial
curvature perturbations (left plot) and tensor-to-scalar ratio
(right plot) at the first horizon crossing that correspond to
values $n_{\mathcal{S}}=[0.9607,0.9691]$ and $r<0.064$
respectively and also satisfy the approximation (\ref{approxphi}),
for $N=60$. }\label{nsrpl}
\end{figure}
The second slow-roll index also takes a simple form,
$\epsilon_2=\frac{4 \mathcal{V}_0}{3 \kappa ^4 M^2 \varphi ^2}$,
while $\epsilon_3=\frac{4 \mathcal{V}_0 \left(3 \kappa ^2 M^2+2
\mathcal{V}_0\right)}{-27 \kappa ^4 M^4-6 \kappa ^2 M^2
\mathcal{V}_0 \left(\kappa ^2 \varphi ^2-1\right)+4
\mathcal{V}_0^2}$, and $\epsilon_4$ is more complicated so we do
not quote here. Also, the value of the scalar field at the first
horizon crossing, namely $\varphi_i$, can easily be found using
Eq. (\ref{N}) that gives the $e$-foldings number along with the $\varphi_f$
we obtained earlier, so upon solving (\ref{N}) after performing
the integration, we get
$\varphi_i=\frac{\sqrt{\frac{2}{3}}\sqrt{2\mathcal{V}_0+3M^2\kappa^2+6M^2N\kappa^2}}{M\kappa^2}.$

\begin{figure}
\centering
\includegraphics[width=18pc]{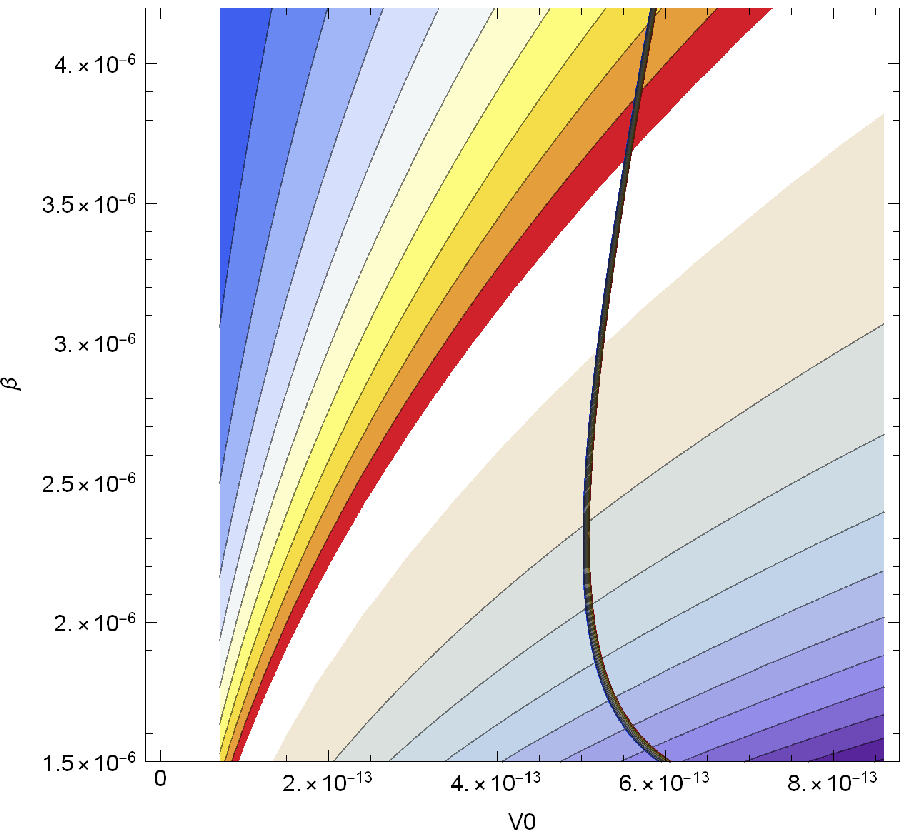}
\includegraphics{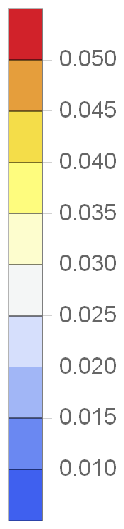}
\includegraphics{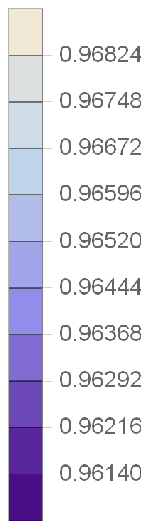}
\caption{Contour plot of the spectral index of primordial
curvature perturbations (lower contours), the factor
$\frac{\kappa^2 \dot \varphi^2}{M^2}$ (upper contours) and the
amplitude of the scalar curvature perturbations (dark line) at the
first horizon crossing, for $N=68$, indicating that a simultaneous
satisfaction of their constraints cannot be achieved for any
values of the free parameters when $N>67$.}\label{N68}
\end{figure}

We can now calculate the slow-roll indices, the spectral index,
the tensor-to-scalar ratio, the tensor spectral index and the
amplitude of the scalar curvature perturbations at the first
horizon crossing by setting $\varphi=\varphi_i$. The final
relations are omitted for brevity. We generally assume that
$M=\beta/\kappa$, where $\beta$ is some dimensionless parameter,
therefore the model has three dimensionless free parameters:
$\beta$, $\mathcal{V}_0$ and the $e$-foldings number, $N$, hence
it is a three parameter model. What we generally expect is that
$\mathcal{V}_0$ will be differently constrained in the
$\mathcal{R}^2$-corrected theory than it is constrained in the
simple quadratic inflation model. We shall examine for which
values of these parameters this model provides a viable
inflationary phenomenology. According to the latest Planck data,
the spectral index and the tensor-to-scalar ratio are constrained
as following \cite{Planck:2018jri},
\begin{equation}\label{constraints}
    n_{\mathcal{S}}=0.9649 \pm 0.0042 \ ,  \ r<0.064 .
\end{equation}
Furthermore, the power spectrum of the scalar perturbations should be constrained to the values,
\begin{equation}\label{psconstraints}
    \mathcal{P}_{\zeta}(k)=2.19\pm0.02\times 10^{-9},
\end{equation}
for a viable theory. As it occurs, the amplitude of the scalar
perturbations basically constrains severely the final model, and
our analysis of the parameter space showed that it is compatible
with the Planck constraints \cite{Planck:2018jri} when the values
for the free parameters are of order $\mathcal{V}_0 \sim
\mathcal{O}(10^{-13})$ and $\beta \sim \mathcal{O}(10^{-6})$. For
example, if we set $\mathcal{V}_0=9.37 \times 10^{-13}$,
$\beta=6.8 \times 10^{-6}$ for $N=60$ we get,
\begin{equation}\label{nsrexample}
   n_{\mathcal{S}}=0.96611,\,\,\,
   r=0.063968,\,\,\,n_{\mathcal{T}}=-0.00799592\, ,
\end{equation}
so the phenomenology is compatible with the latest Planck data
\cite{Planck:2018jri} and the tensor spectrum is red-tilted. As
for the amplitude of the scalar perturbations, we have,
\begin{equation}\label{psexample}
   \mathcal{P}_{\zeta}(k)= 2.19216 \times 10^{-9},
\end{equation}
so this quantity too is compatible with the Planck constraints. We
remind that we should constantly check the validity of the
approximations (\ref{approxH}), (\ref{approxphi}) for the values
of the free parameters that we select. The model is proven to be
compatible with the latest Planck constraints, for a wide range of
the free parameters $\mathcal{V}_0$ and $\beta$.
\begin{figure}
\centering
\includegraphics[width=20pc]{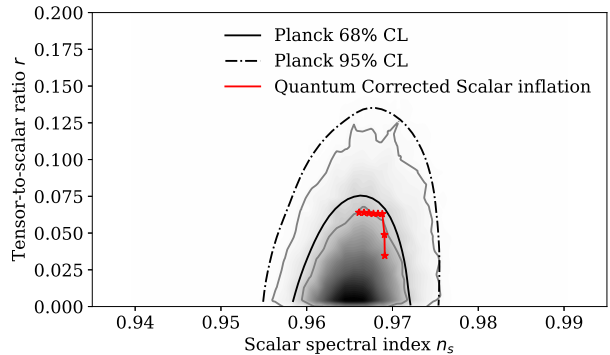}
\caption{Confrontation of the $\mathcal{R}^2$ corrected quadratic
inflation gravity model (red curve) with the 2018 Planck
constraints for $N$ varying in the range $N=[60,67]$ and a
selection of values for the free parameters that result in values
of the observational indices in accordance with the latest Planck
constraints, respecting the approximations following the
theory.}\label{likcurv}
\end{figure}
Indeed, the values used in our example are not the only values
that can make the observational indices simultaneously compatible
with the Planck constraints, as can be seen in Fig.\ref{nsrpl},
where we present the contour plots of the spectral index and the
tensor-to-scalar ratio for a range of values of the free
parameters,that correspond $n_{\mathcal{S}}=[0.9607,0.9691]$ and
$r<0.064$ and also obey the approximations (\ref{approxH}) and
(\ref{approxphi}), that hold a key role in our analysis. Moreover,
in Fig.\ref{likcurv} we also present the 2018 Planck likelihood
curves and one can notice that the $\mathcal{R}^2$-corrected
quadratic inflation theory is well fitted within those likelihood
curves. The resulting line is composed by a variety of points for
different values of the free parameters, $\mathcal{V}_0, \beta$
and for number of $e$-foldings in the range $N=[60,67]$ such, that
the values of the observational indices calculated satisfy the
Planck constraints (\ref{constraints}), the amplitude of the
scalar perturbations falls within the range of
(\ref{psconstraints}) and the approximations
(\ref{approxH}),(\ref{approxphi}) hold true. Generally, the
approximation (\ref{approxH}) holds true for a very wide range of
values of $\mathcal{V}_0$ and $\beta$, however we cannot say the
same for (\ref{approxphi}). The latter approximation is harder to
satisfy, thus it impacts the allowed values of the free parameters
resulting in further limitations to the values of
$n_{\mathcal{S}}$ and $r$. The reason that we stopped adding
points at $N=67$ is that for greater values of $N$ there are no
values of $\mathcal{V}_0$ and $\beta$ that result in the spectral
index taking values within $n_{\mathcal{S}}=[0.9607,0.9691]$ and
in simultaneous validity of the approximation (\ref{approxphi})
along with the amplitude of the scalar perturbations satisfying
(\ref{psconstraints}). This behavior is depicted in Fig.
\ref{N68}, where the upper contours of the plot represent the
values of $\frac{\kappa^2 \dot \varphi^2}{M^2}$, the lower
contours the values of $n_{\mathcal{S}}$ that satisfy the Planck
constraints and the thin textured line the accepted values for the
amplitude of the scalar perturbations. It is obvious that there is
no overlapping of these plots for any values of the free
parameters, a characteristic that becomes more intense as $N$
increases beyond $N>67$.
Summarizing, the $\mathcal{R}^2$ quantum corrected quadratic
inflation model yields an inflationary phenomenology which is
compatible with the latest Planck constraints, in contrast to the
simple quadratic inflation model, which is incompatible with the
Planck data. In principle more examples of this sort can be
studied, but the analysis is more or less the same, so we do not
go in details on extra models.

\section{Conclusions}

In this work we studied a generalized $f(\mathcal{R},\varphi)$
gravity framework, in which a canonical scalar field theory is
considered in the presence of string theory originating higher
order curvature quantum correction terms. We considered for
simplicity an $\mathcal{R}^2$ term and we investigated in detail
the effect of the $\mathcal{R}^2$ correction term on the equations
of motion of scalar field theory at leading order. After deriving
the field equations of the combined $\mathcal{R}^2$-corrected
scalar field theory, we presented the final form of the slow-roll
indices for this theory. As we showed, the latter acquire quite
elegant final forms, in which the leading order effect of the
$\mathcal{R}^2$ quantum corrections is directly seen in each one
of them. The resulting theoretical framework was tested with
respect to its inflationary aspects. For a test model we chose the
quadratic inflation model, which is a non-viable single scalar
field model when it is considered by itself. As we showed, the
$\mathcal{R}^2$-corrected quadratic inflation model yields a
viable inflationary phenomenology. With regard to the inflationary
phenomenology, we mainly focused on the spectral index of the
primordial scalar curvature perturbations, the tensor-to-scalar
ratio, the tensor spectral index and the amplitude of the scalar
perturbations. One notable feature is the fact that the
multiplicative parameter of the potential, which we denoted as
$\mathcal{V}_0$, is constrained differently in comparison to the
simple quadratic inflation model. Indeed, the final form of the
amplitude of the scalar perturbations in the
$\mathcal{R}^2$-corrected quadratic inflation theory is different
from that of the simple quadratic inflation theory, thus
$\mathcal{V}_0$ can take different values compared to the simple
quadratic inflation case. The next step toward the research line
we adopted in this paper, is to considered combinations of quantum
corrections in the simple scalar field theory. The case of an
$\mathcal{R}^2$-corrected Einstein-Gauss-Bonnet theory was
considered in Ref. \cite{Odintsov:2020ilr}, and it was shown that
a viable phenomenology can be obtained for these models, for both
minimal and non-minimally coupled scalar field models. What has
not been studied yet in the literature is considering
$\mathcal{R}^3$ corrections, which basically originate from
order-six higher curvature derivatives in the quantum corrected
scalar field action (\ref{quantumaction}). We will address this
issue in the near future.

Finally, let us comment that the whole point of this work was to
simply introduce an alternative to the standard scalar field
inflation including $R^2$ corrections in the scalar field
Lagrangian. The potential we chose for checking our formalism is
an $\phi^2$ potential, which without the $R^2$ term is not
compatible with the Planck data. But with our formalism we showed
that the model is viable, if $R^2$ corrections are considered. So
our intention was not to compare our model with the standard $R^2$
model, but to see what happens if one uses both scalar fields and
higher order curvature terms in the inflationary Lagrangian. If
those are considered together in the Jordan frame, we showed how
such a framework can lead to sensible results. In Einstein frame
descriptions this is not possible though because dealing
analytically with two scalar fields is impossible to work easily.


\begin{thebibliography}{99}







\bibitem{inflation1}
 A.~D.~Linde,
 Lect.\ Notes Phys.\ {\bf 738} (2008) 1
 [arXiv:0705.0164 [hep-th]].

\bibitem{inflation2} D.~S.~Gorbunov and V.~A.~Rubakov,
``Introduction to the theory of the early universe: Cosmological
perturbations and inflationary theory,'' Hackensack, USA: World
Scientific (2011) 489 p;
%


\bibitem{inflation3}A.~Linde,
arXiv:1402.0526 [hep-th];


\bibitem{inflation4}D.~H.~Lyth and A.~Riotto,
Phys.\ Rept.\  {\bf 314} (1999) 1 [hep-ph/9807278].



\bibitem{CMB-S4:2016ple}
K.~N.~Abazajian \textit{et al.} [CMB-S4],
[arXiv:1610.02743 [astro-ph.CO]].



\bibitem{SimonsObservatory:2019qwx}
M.~H.~Abitbol \textit{et al.} [Simons Observatory],
Bull. Am. Astron. Soc. \textbf{51} (2019), 147 [arXiv:1907.08284
[astro-ph.IM]].







\bibitem{Kamionkowski:2015yta}
  M.~Kamionkowski and E.~D.~Kovetz,
  Ann.\ Rev.\ Astron.\ Astrophys.\  {\bf 54} (2016) 227
  doi:10.1146/annurev-astro-081915-023433
  [arXiv:1510.06042 [astro-ph.CO]].






\bibitem{Hild:2010id}
S.~Hild, M.~Abernathy, F.~Acernese, P.~Amaro-Seoane, N.~Andersson,
K.~Arun, F.~Barone, B.~Barr, M.~Barsuglia and M.~Beker, \textit{et
al.}
Class. Quant. Grav. \textbf{28} (2011), 094013
doi:10.1088/0264-9381/28/9/094013 [arXiv:1012.0908 [gr-qc]].




\bibitem{Baker:2019nia}
J.~Baker, J.~Bellovary, P.~L.~Bender, E.~Berti, R.~Caldwell,
J.~Camp, J.~W.~Conklin, N.~Cornish, C.~Cutler and R.~DeRosa,
\textit{et al.}
[arXiv:1907.06482 [astro-ph.IM]].


\bibitem{Smith:2019wny}
T.~L.~Smith and R.~Caldwell,
Phys. Rev. D \textbf{100} (2019) no.10, 104055
doi:10.1103/PhysRevD.100.104055 [arXiv:1908.00546 [astro-ph.CO]].


\bibitem{Crowder:2005nr}
J.~Crowder and N.~J.~Cornish,
Phys. Rev. D \textbf{72} (2005), 083005
doi:10.1103/PhysRevD.72.083005 [arXiv:gr-qc/0506015 [gr-qc]].


\bibitem{Smith:2016jqs}
T.~L.~Smith and R.~Caldwell,
Phys. Rev. D \textbf{95} (2017) no.4, 044036
doi:10.1103/PhysRevD.95.044036 [arXiv:1609.05901 [gr-qc]].



\bibitem{Seto:2001qf}
N.~Seto, S.~Kawamura and T.~Nakamura,
Phys. Rev. Lett. \textbf{87} (2001), 221103
doi:10.1103/PhysRevLett.87.221103 [arXiv:astro-ph/0108011
[astro-ph]].


\bibitem{Kawamura:2020pcg}
S.~Kawamura, M.~Ando, N.~Seto, S.~Sato, M.~Musha, I.~Kawano,
J.~Yokoyama, T.~Tanaka, K.~Ioka and T.~Akutsu, \textit{et al.}
[arXiv:2006.13545 [gr-qc]].



\bibitem{Bull:2018lat}
A.~Weltman, P.~Bull, S.~Camera, K.~Kelley, H.~Padmanabhan,
J.~Pritchard, A.~Raccanelli, S.~Riemer-S\o{}rensen, L.~Shao and
S.~Andrianomena, \textit{et al.}
Publ. Astron. Soc. Austral. \textbf{37} (2020), e002
doi:10.1017/pasa.2019.42 [arXiv:1810.02680 [astro-ph.CO]].




\bibitem{Codello:2015mba}
A.~Codello and R.~K.~Jain,
Class. Quant. Grav. \textbf{33} (2016) no.22, 225006
doi:10.1088/0264-9381/33/22/225006 [arXiv:1507.06308 [gr-qc]].






\bibitem{reviews1}
 S.~Nojiri, S.~D.~Odintsov and V.~K.~Oikonomou,
  Phys.\ Rept.\  {\bf 692} (2017) 1
  [arXiv:1705.11098 [gr-qc]].

\bibitem{reviews2}


 S. Capozziello, M. De Laurentis,
   Phys.\ Rept.\  {\bf 509}, 167 (2011);\\
 V.~Faraoni and S.~Capozziello,
  Fundam.\ Theor.\ Phys.\  {\bf 170} (2010).



\bibitem{reviews3}
S. Nojiri, S.D. Odintsov,
  eConf {\bf C0602061}, 06 (2006)
  [Int.\ J.\ Geom.\ Meth.\ Mod.\ Phys.\  {\bf 4}, 115 (2007)].


   \bibitem{reviews4}

S. Nojiri, S.D. Odintsov,
   Phys.\ Rept.\  {\bf 505}, 59 (2011);




\bibitem{reviews5}

A.~de la Cruz-Dombriz and D.~Saez-Gomez,
  Entropy {\bf 14} (2012) 1717
  [arXiv:1207.2663 [gr-qc]].

\bibitem{reviews6}

G.~J.~Olmo,
  Int.\ J.\ Mod.\ Phys.\ D {\bf 20} (2011) 413
  [arXiv:1101.3864 [gr-qc]].











\bibitem{Ema:2017rqn}
Y.~Ema,
Phys. Lett. B \textbf{770} (2017), 403-411
doi:10.1016/j.physletb.2017.04.060 [arXiv:1701.07665 [hep-ph]].



\bibitem{Ema:2020evi}
Y.~Ema, K.~Mukaida and J.~Van De Vis,
JHEP \textbf{02} (2021), 109 doi:10.1007/JHEP02(2021)109
[arXiv:2008.01096 [hep-ph]].



\bibitem{Ivanov:2021ily}
V.~R.~Ivanov and S.~Y.~Vernov,
[arXiv:2108.10276 [gr-qc]].




\bibitem{Gottlober:1993hp}
S.~Gottlober, J.~P.~Mucket and A.~A.~Starobinsky,
Astrophys. J. \textbf{434} (1994), 417-423 doi:10.1086/174743
[arXiv:astro-ph/9309049 [astro-ph]].





\bibitem{delaCruz-Dombriz:2016bjj}
A.~de la Cruz-Dombriz, E.~Elizalde, S.~D.~Odintsov and
D.~S\'aez-G\'omez,
JCAP \textbf{05} (2016), 060 doi:10.1088/1475-7516/2016/05/060
[arXiv:1603.05537 [gr-qc]].



\bibitem{Enckell:2018uic}
V.~M.~Enckell, K.~Enqvist, S.~Rasanen and L.~P.~Wahlman,
JCAP \textbf{01} (2020), 041 doi:10.1088/1475-7516/2020/01/041
[arXiv:1812.08754 [astro-ph.CO]].



\bibitem{Karam:2018mft}
A.~Karam, T.~Pappas and K.~Tamvakis,
JCAP \textbf{02} (2019), 006 doi:10.1088/1475-7516/2019/02/006
[arXiv:1810.12884 [gr-qc]].




\bibitem{Kubo:2020fdd}
J.~Kubo, J.~Kuntz, M.~Lindner, J.~Rezacek, P.~Saake and
A.~Trautner,
JHEP \textbf{08} (2021), 016 doi:10.1007/JHEP08(2021)016
[arXiv:2012.09706 [hep-ph]].



\bibitem{Gorbunov:2018llf}
D.~Gorbunov and A.~Tokareva,
Phys. Lett. B \textbf{788} (2019), 37-41
doi:10.1016/j.physletb.2018.11.015 [arXiv:1807.02392 [hep-ph]].



\bibitem{Calmet:2016fsr}
X.~Calmet and I.~Kuntz,
Eur. Phys. J. C \textbf{76} (2016) no.5, 289
doi:10.1140/epjc/s10052-016-4136-3 [arXiv:1605.02236 [hep-th]].



\bibitem{Oikonomou:2021msx}
V.~K.~Oikonomou,
Annals Phys. \textbf{432} (2021), 168576
doi:10.1016/j.aop.2021.168576 [arXiv:2108.04050 [gr-qc]].





\bibitem{Odintsov:2021kup}
S.~D.~Odintsov, V.~K.~Oikonomou and F.~P.~Fronimos,
[arXiv:2108.11231 [gr-qc]].




\bibitem{Odintsov:2017hbk}
S.~D.~Odintsov, V.~K.~Oikonomou and L.~Sebastiani,
Nucl. Phys. B \textbf{923} (2017), 608-632
doi:10.1016/j.nuclphysb.2017.08.018 [arXiv:1708.08346 [gr-qc]].






\bibitem{Starobinsky:1980te}
A.~A.~Starobinsky,
Phys. Lett. B \textbf{91} (1980), 99-102
doi:10.1016/0370-2693(80)90670-X



\bibitem{Bezrukov:2007ep}
F.~L.~Bezrukov and M.~Shaposhnikov,
Phys. Lett. B \textbf{659} (2008), 703-706
doi:10.1016/j.physletb.2007.11.072 [arXiv:0710.3755 [hep-th]].




\bibitem{Appleby:2009uf}
S.~A.~Appleby, R.~A.~Battye and A.~A.~Starobinsky,
JCAP \textbf{06} (2010), 005 doi:10.1088/1475-7516/2010/06/005
[arXiv:0909.1737 [astro-ph.CO]].





\bibitem{Hwang:2005hb}
  J.~c.~Hwang and H.~Noh,
  Phys.\ Rev.\ D {\bf 71} (2005) 063536
  doi:10.1103/PhysRevD.71.063536
  [gr-qc/0412126].





\bibitem{Planck:2018jri}
Y.~Akrami \textit{et al.} [Planck],
Astron. Astrophys. \textbf{641} (2020), A10
doi:10.1051/0004-6361/201833887 [arXiv:1807.06211 [astro-ph.CO]].




\bibitem{Odintsov:2020ilr}
S.~D.~Odintsov, V.~K.~Oikonomou and F.~P.~Fronimos,
Annals Phys. \textbf{424} (2021), 168359
doi:10.1016/j.aop.2020.168359 [arXiv:2011.08680 [gr-qc]].








\end{thebibliography}
\end{document}